\documentstyle[aps,preprint]{revtex}
\begin{document}
\draft
\title{Entropic Sampling and Natural Selection\\
       in Biological Evolution}
\author{M.Y. Choi, H.Y. Lee, D. Kim, and S.H. Park}
\address{
     Department of Physics and Center for Theoretical Physics\\
     Seoul National University\\
     Seoul 151-742, Korea}

\maketitle

\begin{abstract}
With a view to connecting random mutation on the molecular level to
punctuated equilibrium behavior on the phenotype level,
we propose a new model for biological evolution,
which incorporates random mutation and natural selection.
In this scheme the system evolves continuously into new configurations,
yielding non-stationary behavior of the total fitness.
Further, both the waiting time distribution of species and the avalanche
size distribution display power-law behaviors with exponents close
to two, which are consistent with the fossil data. 
These features are rather robust, indicating the key role of entropy.
\end{abstract}

\thispagestyle{empty}
\pagebreak
\narrowtext

The idea of punctuated equilibrium, based on the new interpretation of
the fossil record, holds that biological evolution proceeds 
not at a steady pace but in an intermittent manner \cite{gould}. 
Simple models attempting to describe such phenomena
usually employ the extremal dynamics, which evolves the system 
by sequentially updating or mutating the species 
with the globally minimum value of fitness~\cite{bak,Sneppen}.
This indeed drives the system into a self-organized critical state,
characterized by the power-law behaviors,
and such extremal dynamics is regarded as an essential
ingredient to achieve the self-organized criticality.
In reality, however, there is no reason why mutation necessarily occurs in the
species with the minimum fitness.
On the contrary, on the molecular level most evolutionary changes
and most of variability within a species are believed to be caused 
not by selection but by random drift of mutant genes 
which are selectively equivalent \cite{dob}.
In contrast to such randomness on the molecular level,
the fossil data exhibit critical behavior on the phenotypic level:
For example, it is known that the number $M_t$ of genera with lifetime $t$
displays the power-law behavior, $M_t \sim t^{-\alpha}$ 
with $\alpha \approx 2$~\cite{raupsep}.
Here, the longer a genus exists in the ecosystem,
the more sub-genera should be generated; it is thus expected that
the number of sub-families is proportional to the length of the period
of existence~\cite{adami,willis}. 
Indeed recent analysis of fossil records has revealed such structure
in the taxonomic system
that the number $M_n$ of taxa having $n$ sub-taxa
follows essentially the same power-law distribution:
$M_n \sim n^{-\alpha'}$ with 
$\alpha' \approx 1.84$ to $2.41$~\cite{burlando}.
These imply that there does not exist a relevant time scale 
characterizing the length of the period during which
a certain species dominates the population.
Furthermore, it is known that a few mass extinction events as well as many
background events (of smaller sizes) have been occurred during last
600 million years~\cite{futuyma},
which also exhibit power-law behavior.  Namely, the
distribution of extinction events of size $s$ is observed to follow
$\tilde{M}(s) \sim s^{-\tau}$ with $1\lesssim \tau \lesssim 2$ \cite{bakpac},
again manifesting scale invariance.

This work attempts to provide a natural link between the behavior
on the molecular level and that on the phenotypic level. 
For this purpose, we propose a new model of evolution, which is more
realistic than existing models, and show how the power-law behavior 
emerges on the phenotype level.
Our model incorporates the {\em random mutation} on the molecular level
and introduces an additional step determining acceptance or rejection 
of a particular mutation.  The latter mimics the {\em natural selection}
process, connecting the molecular and phenotypic levels.
Here the entropy of the ecosystem plays a key role:
When information is transferred from the environment to the species 
via mutation, the information-theoretic entropy of the species drops.
Although the subsequent (random) mutation may restore the previous level 
of the entropy by reestablishing disorder,
the mutation is on the average oriented to lower the entropy during
the process of evolution \cite{adami}. 
This corresponds
to transferring information from the environment to the species
in the view point of information theory, where entropy stands for
missing (negative) information. 
We consider various forms of the fitness function, 
and measure the waiting time distributions of the species having maximum 
fitness values as well as of a typical species.  They all show power-law
behaviors with exponents close to two.  
We also investigate the size distributions of ``avalanches'', i.e.,
series of mutations triggered by single starting mutations,
again to obtain power-law behavior, which is consistent with
the observed distribution of extinction events.

Consider an ecosystem consisting of $N$ species, 
in interaction with the environment.
The configuration of the ecosystem is described by $x \equiv \{x_i\}$, where
$x_i$ represents the configuration of the $i$th species.
Usually, $x_i$ is taken to be a number between zero and unity 
($0\leq x_i \leq 1$).
We assume that the fitness of the $i$th species is modeled by a real function
$f_i (x)$, which may in general depend upon the 
configuration of neighboring species as well as its own configuration $x_i$.
The total fitness $F(x)$ is then defined to be the sum of the fitness values
of all species in the system:
\begin{equation}
  F(x)=\sum_{i=1}^N f_i(x).
\end{equation}
Accordingly, the phase space 
volume (or the number of accessible states) $\Omega(F)$ 
for given value of the total fitness $F$ reads
\begin{equation}
 \Omega (F) = \prod_{j=1}^N \int_0^1 dx_j \,
               \delta [F-\sum_{i=1}^N f_i(x)], \label{omega}
\end{equation}
the logarithm of which gives the entropy of the ecosystem.

In an ecosystem mutation is an exception to the regularity of 
the process of DNA replication, 
which normally involves precise copying of the hereditary
information encoded in nucleotide sequences.  
We thus choose randomly a single species in the system 
and mutate the chosen species only, since there is no obvious reason
that mutations occur in the nearest neighboring species simultaneously 
\cite{chau}.
The natural selection via which the mutation
is accepted or rejected is determined by the entropy in our scheme. 
We assume that the change in the total entropy $S_{tot}\equiv S+S_0$,
where $S$ and $S_0$ are the entropy of the ecosystem and that of
the environment, respectively, is negligible during the information 
transfer from the environment to the ecosystem (i.e., the entropy flow
from the ecosystem to the environment).
Namely, the information transfer is assumed to be almost reversible.
(Relaxation of the reversibility assumption will be discussed later.) 
The probability $P(x)$ of finding the system in configuration $x$ 
in the stationary state is simply proportional to the
number $\Omega_0$ of accessible states for the environment:
\begin{equation}
 P(x) \propto \Omega_0 = \exp[S_0] = C \exp[-S(F)], \label{p}
\end{equation}
where $C\equiv \exp[S_{tot}]$ is a constant and the entropy of the ecosystem,
given by the logarithm of the number of accessible states for the system:
$S(F)=\ln\Omega(F)$,
depends on the configuration through the total fitness $F\equiv F(x)$.
The probability distribution given by Eq.~(\ref{p}) can be attained 
dynamically by imposing the detailed balance condition
\begin{equation}
 {{W(x\rightarrow x')}\over {W(x' \rightarrow x)}} = 
  \exp [S(F)-S(F')],  \label{W}
\end{equation}
where the transition probability $W(x{\rightarrow} x')$ describes the
mutation from configuration $x$ to configuration $x'$ and $F' \equiv F(x')$
denotes the total fitness of the ecosystem in configuration $x'$. 
We thus compute the entropy change $\Delta S \equiv S(F')-S(F)$ 
during the process of mutation $x \rightarrow x'$,
and accept the mutation when $\Delta S < 0$; for $\Delta S >0$, 
the mutation is accepted with probability $\exp (-\Delta S)$.
In consequence, the entropy of the ecosystem tends to decrease while
that of the environment tends to increase, since the total entropy remains 
approximately constant (and cannot decrease in particular).
Thus the entropy is directed to flow from the ecosystem to the environment.

Interestingly, the dynamics described by Eq.~(\ref{W})
precisely corresponds to the entropic sampling algorithm, which
was successfully applied to the traveling salesman problem \cite{lee}.
In this algorithm, the entropy of the ecosystem is estimated as follows:
Initially the entropy $S(F)$ is set equal to zero for all values of the
total fitness $F$.
We then obtain the histogram $H(F)$ of the total fitness for a short run,
which gives new estimation of $S(F)$:
\begin{equation}
 S(F)= 
 \left\{
  \begin{array}{lr}
  S(F)& \mbox{for $H(F)=0$},\\
  S(F)+\ln H(F)& \mbox{otherwise}.
  \end{array}
 \right.
\end{equation}
This yields the entropy increasing as the run proceeds.  Once the stationary
state is reached, however, the increase due to more runs 
gives merely an additive constant (independent of $F$), which is obviously
irrelevant. 
Note also that this sampling has characteristics of
frequency-dependent selection~\cite{futuyma} and yields uniform distribution 
of the total fitness over the entire space, which appears natural in
a real ecosystem.
It assists the system to
escape from a local maximum in the total fitness space 
and fall into a new local maximum, which
corresponds to a metastable phenotype of the ecosystem.
In this manner the ecosystem is allowed to evolve into a new configuration
by gathering information from the environment 
(i.e., by reducing its entropy). 
In particular, the additional step determining the acceptance/rejection 
of mutation naturally leads to correlations between the configuration
before the mutation and that after it.
%
For the simplest Bak-Sneppen (BS) type fitness~\cite{bak}
\begin{equation}
 f_i = x_i, \label{b}
\end{equation}
which 
allows the explicit evaluation of Eq.~(\ref{omega}),
the above algorithm indeed yields the entropy
in perfect coincidence with that obtained from Eq.~(\ref{omega})
(up to an irrelevant additive constant). 
This demonstrates the accuracy of the entropic sampling although
other forms of the fitness function in general 
do not allow analytic calculation to be compared.

In most of the existing models, the total fitness remains constant 
once the stationary state is reached.  This implies that the
species in the present time and those in the past have the same average 
adaptive power to the environment, which does not appear to be 
true in reality \cite{chau}.
In contrast, the new model, which takes into account the role of entropy, 
leads to non-stationary behavior of the total fitness.
In the entropic sampling algorithm, which is adopted in the new model, 
the mutation is encouraged to decrease the entropy of the ecosystem and
to transfer information from the environment to the species.
Here the entropy of the system increases if the system stays long in 
configurations with the same value of $F$. 
Accordingly, the probability for the mutation making
the system escape from a region of the same $F$
is higher than that for the mutation having the system stay 
in that region, producing non-stationary behavior. 
In this way the system evolves continuously into new configurations
in the whole fitness landscape instead of staying at a single local maximum,
which reflects the interactions with the changing environment.

To compare with the fossil data, which give the exponent
$\alpha$ close to two in the power-law behaviors
of $M_t$ and $M_n$,
we have chosen a species at random
and measured the distribution of the interval between two consecutive
mutations.
The resulting (average) waiting time distribution $D(t)$ of the species,
measured during the time
$10^4$ to $5 \times 10^8$ and averaged over five independent
runs with different initial configurations, is shown in Fig.~1.  It
indeed exhibits power-law behavior~\cite{spread}: 
$D \sim t^{-\nu}$ for $t \geq 3$ with
the exponent $\nu =2.07 \pm 0.02$, 
which is consistent with the frequency distribution of taxa
observed in the fossil data; this is
significantly larger than the value $1.57$ obtained 
in the conventional model with the extremal dynamics~\cite{maslov}.
In reality the contributions of various species to the fossil record are not
all equal and those species which have the maximum fitness values
should dominate the record.
This implies the relative importance of the species having the maximum 
fitness values in the comparison with the fossil data~\cite{adami}.
We have thus watched the species having maximum fitness
and measured the waiting time distribution of such species 
during the same time interval, performing ten independent 
runs with different initial configurations.
The resulting waiting time distribution of the species having maximum
fitness again exhibits power-law behavior with
the exponent $\nu =2.11 \pm 0.04$.  The overall features are similar to 
those of the average distribution shown in Fig.~1, but the power-law nature
is more clear and accurate, yielding an almost perfect fit for $t\geq 3$.
The almost identical behavior apparently indicates the
validity of the results regardless of the (unknown) relative importance
in the contributions to the fossil data.

To check whether these features are sensitive to the specific choice of
the fitness function, we have also considered other types of
the fitness function, e.g.,
\begin{equation}
 f_i=\mathop{{\sum_j}'}{(x_i-x_j-A_{ij})}^2, \label{g}
\end{equation}
where the prime restricts the summation to 
nearest-neighboring species of $i$ on a square lattice
and $A_{ij}$ has been introduced to accommodate the possibility of 
``frustration'' in the interactions between species.
Since the exact form of interactions between species is not available,
$A_{ij}$'s are taken to be quenched random variables, taking the values
between ${-}5$ and $5$.
Unlike in Eq.~(\ref{b}), the fitness in Eq.~(\ref{g}) 
is determined not only by its own configuration but also by the 
configurations of neighboring species,
and can change due to the mutation in 
neighboring species, which is presumably more realistic.
The fitness given by Eq.~(\ref{g}) has been investigated in a system of $16^2$
species arranged to form a square array with linear size $L=16$. 
The same entropic sampling algorithm again yields
the waiting time distribution of the species with maximum fitness values, 
which is fitted to the power-law behavior with exponent $\nu=2.08 \pm 0.07$, 
as shown in Fig.~2. 
The distribution has been measured during the time $10^4$ to 
$2\times 10^7$,
and five independent runs have been performed with different initial
configurations and realizations of $A_{ij}$'s \cite{spread}. 
The average waiting time distribution of a typical species has also 
been obtained, revealing essentially the same power-law behavior except for
the exponent $\nu=2.12 \pm 0.09$. 
We have also checked other distribution of $A_{ij}$'s,
and found that the overall behavior still does not change qualitatively.
This apparently suggests that the general behavior is rather insensitive
to the specific form of the fitness function, giving
support to the use of a simple model without knowing the
precise dynamics of the evolution. 

When a mutation is accepted, the fitness of some species and
consequently, the total fitness change.  This in turn leads to a new value
of the entropy, which determines the transition probability to another 
configuration, i.e., the mutation probability.
Therefore the acceptance/rejection of a mutation depends on that of the 
previous mutation, and
the resulting correlations between the evolving configurations may be
characterized by the distribution of ``avalanches'',
which stand for the series of (accepted)
mutations triggered by single starting mutations.  Since a mutation
occurring on a species implies change of the species into another,
avalanches correspond to extinctions of species,
for which fossil data give power-law behavior in the size
distribution, with the exponent $\tau$ between $1$ and $2$~\cite{bakpac}.
To accommodate the large interval in the fossil analysis
compared with the inverse of the mutation rate in nature,
We have regarded the mutation accepted within $\Delta$ trials
following a mutation as constituting an avalanche, and
measured the distribution $\tilde{D}(s)$ of the avalanche events with
size $s$ for various values of the interval $\Delta$.
It is found that $\tilde{D}(s)$ indeed displays power-law behavior:
$\tilde{D}(s) \sim s^{-\mu}$ with the exponent $\mu \approx 2$,
unless $\Delta$ is too small.   
Figure 3 shows the behavior of the avalanche size distribution for
$\Delta =10$, computed up to the time $10^6$ in a system of 1000 species
(with the BS type fitness). 
The exponent is given by $\mu = 1.98 \pm 0.01$, which is consistent with
the analysis of the fossil data.  We have also considered larger values of 
the interval, up to $\Delta =20$, and obtained almost the same 
power-law behavior with $\mu$ hardly changing; we thus believe that
such power-law behavior is robust and persistent even for realistically
large values of $\Delta$.

We now consider the 
effects of irreversible information transfer 
in the natural selection process.
In this case, the total entropy $S_{tot}$ does not remain constant but
increases during the information transfer,
via which the ecosystem also increases its fitness $F$.
It is thus plausible to assume that
$S_{tot}$ is an increasing function of $F$:
$S_{tot}(F) \approx S_{tot}(F_0) + \beta (F-F_0)$, for some reference 
value $F_0$, where $\beta \equiv \partial S_{tot}/\partial F|_{F=F_0} > 0$.
This leads to the probability for the ecosystem, $
 P(x) 
 \propto \exp[\beta F -S(F)]$,
which can be attained by 
the following algorithm:
In determining acceptance or rejection of a mutation, we
consider the change in the total fitness, $\Delta F$, 
in addition to the entropy change $\Delta S$,
and select the new configuration according to the probability
$\mbox{min}[1, e^{-\Delta S +\beta \Delta F}]$, 
where $\beta$ measures the relative importance of the total fitness.
This algorithm, where the available range of the total
fitness is controlled by the term $e^{\beta \Delta F}$,
yields qualitatively the same results:
The system still displays the 
non-stationary behavior of the
total fitness and the power-law behavior of the waiting time distribution 
with a similar value of the exponent.  
For example, for the fitness function given by Eq.~(\ref{g}), we obtain 
$\nu=2.03\pm 0.05$ unless 
$\beta$ is too large. 
On the other hand, in the absence of the entropy term, 
the generalized algorithm reduces to the standard (Metropolis) importance 
sampling algorithm.  This importance sampling, when used 
as an alternative to the entropic sampling in the natural selection process,
fails to yield the power-law behavior; instead it in general gives 
stretched exponential behavior~\cite{palmer} 
which is rather close to the conventional 
exponential behavior, in apparent disagreement with observation.
%

We are grateful to S.Y.  Park for his help in numerical works, 
and acknowledge the partial support from the BSRI
program, Ministry of Education, and from the KOSEF through the SRC program.




\begin{figure}
\caption{Average waiting time distribution of a typical species,
 measured during the time $10^4$ to $5\times 10^8$, in the system
 of 100 species, with the BS-type fitness function. 
 The time has been measured in units of the Monte Carlo steps per species.
 The dashed line corresponds to the least-square fit of the data for
 $t \geq 8$, giving the slope $-2.07 \pm 0.02$.}
\end{figure}

\begin{figure}
\caption{Waiting time distribution of the species having maximum fitness,
 measured during the time $10^4$ to $2\times 10^7$, in the system of 
 $16^2$ species with the Gaussian fitness function.
 The dashed line represents the least-square fit of the data for $t \geq 3$,
 with the slope $-2.08 \pm 0.07$.} 
\end{figure}

\begin{figure}
\caption{Avalanche size distribution 
 measured with the interval $\Delta=10$, in the system of 
 $1000$ species with the BS-type fitness function.
 The dashed line represents the least-square fit of the data,
 with the slope $-1.98 \pm 0.01$.}
\end{figure}





\begin{thebibliography}{99}
\bibitem{spread} Here the data at large times are limited by the sampling 
  time, manifesting the discreteness of the data.  This causes the
  spread of the data, which shifts toward larger $t$ as the sampling time
  is increased.  We have also considered systems of $50$ and $20$ species,
  and found essentially the same behavior.
\end{thebibliography}
\end{document}